\documentclass{vgtc}                          

\ifpdf
  \pdfoutput=1\relax                   
  \usepackage{graphicx}                
  \DeclareGraphicsExtensions{.pdf,.png,.jpg,.jpeg} 
\else
  \usepackage{graphicx}                
  \DeclareGraphicsExtensions{.eps}     
\fi%

\graphicspath{{figures/}{pictures/}{images/}{./}} 

\usepackage{microtype}                 
\PassOptionsToPackage{warn}{textcomp}  
\usepackage{textcomp}                  
\usepackage{mathptmx}                  
\usepackage{times}                     
\usepackage{cite}                      
\usepackage{tabu}                      
\usepackage{booktabs}                  
\usepackage{graphicx} 
\usepackage{subcaption} 
\usepackage[absolute,overlay]{textpos}
\onlineid{0}

\vgtccategory{Research}

\vgtcinsertpkg

\title{Quantitative Analysis of Objects in Prisoner Artworks}

\author{
Thea Christoffersen\textsuperscript{\large *} %
\and Annika Tidemand Jensen\textsuperscript{\large *} %
\and Chris Hall\textsuperscript{\large *} %
\and Christofer Meinecke\textsuperscript{\large †} %
\and Stefan Jänicke\textsuperscript{\large *} %
}
\affiliation{
{\large *} Department of Mathematics and Computer Science, University of Southern Denmark, Odense \\
{\large †} ScaDS.AI Dresden/Leipzig, Leipzig University, Leipzig, Germany
}

\abstract{
Prisoners of Nazi concentration camps created paintings as a means to express their daily life experiences and feelings. Several thousand such paintings exist, but a quantitative analysis of them has not been carried out. We created an extensive dataset of 1,939 Holocaust prisoner artworks, and we employed an object detection framework that found 19,377 objects within these artworks. To support the quantitative and qualitative analysis of the art collection and its objects, we have developed an intuitive and interactive dashboard to promote a deeper engagement with these visual testimonies. The dashboard features various visual interfaces, e.g., a word cloud showing the detected objects and a map of artwork origins, and options for filtering. We presented the interface to domain experts, whose feedback highlights the dashboard’s intuitiveness and potential for both quantitative and qualitative analysis while also providing relevant suggestions for improvement. Our project demonstrates the benefit of digital methods such as machine learning and visual analytics for Holocaust remembrance and educational purposes.
} 

\begin{document}


\firstsection{Introduction}

\maketitle

\begin{textblock*}{8cm}(2cm,24.2cm) 
    \raggedright 
    \small Proceedings of the Prague Visual History and Digital Humanities Conference (PRAVIDCO) 2025
\end{textblock*}

Standing as one of the darkest chapters in human history, the need to preserve and present the memory of the Holocaust is still of vital importance today – not only for remembering and honoring the victims but also for the education of future generations, as the people able to testify to these brutalities are getting fewer and fewer. For years, the testimonies of the Holocaust have been focused primarily on oral and textual testimonies. However, the visual testimonies, though underexplored, also stand as a powerful medium for Holocaust remembrance. During the Holocaust, victims of Nazi persecution, including artists and others, used their creativity to document their experiences and hold on to their humanity under brutal conditions. Creating art in concentration camps, ghettos, or while in hiding was both an act of resistance and a means of survival, despite the severe risks\cite{johnson2023}. These artworks serve as visual testimonies, similar to written accounts, capturing daily struggles, moments of humanity, and resistance against Nazi persecution.

\begin{figure}[ht!]
 \centering 
 \includegraphics[width=\columnwidth]{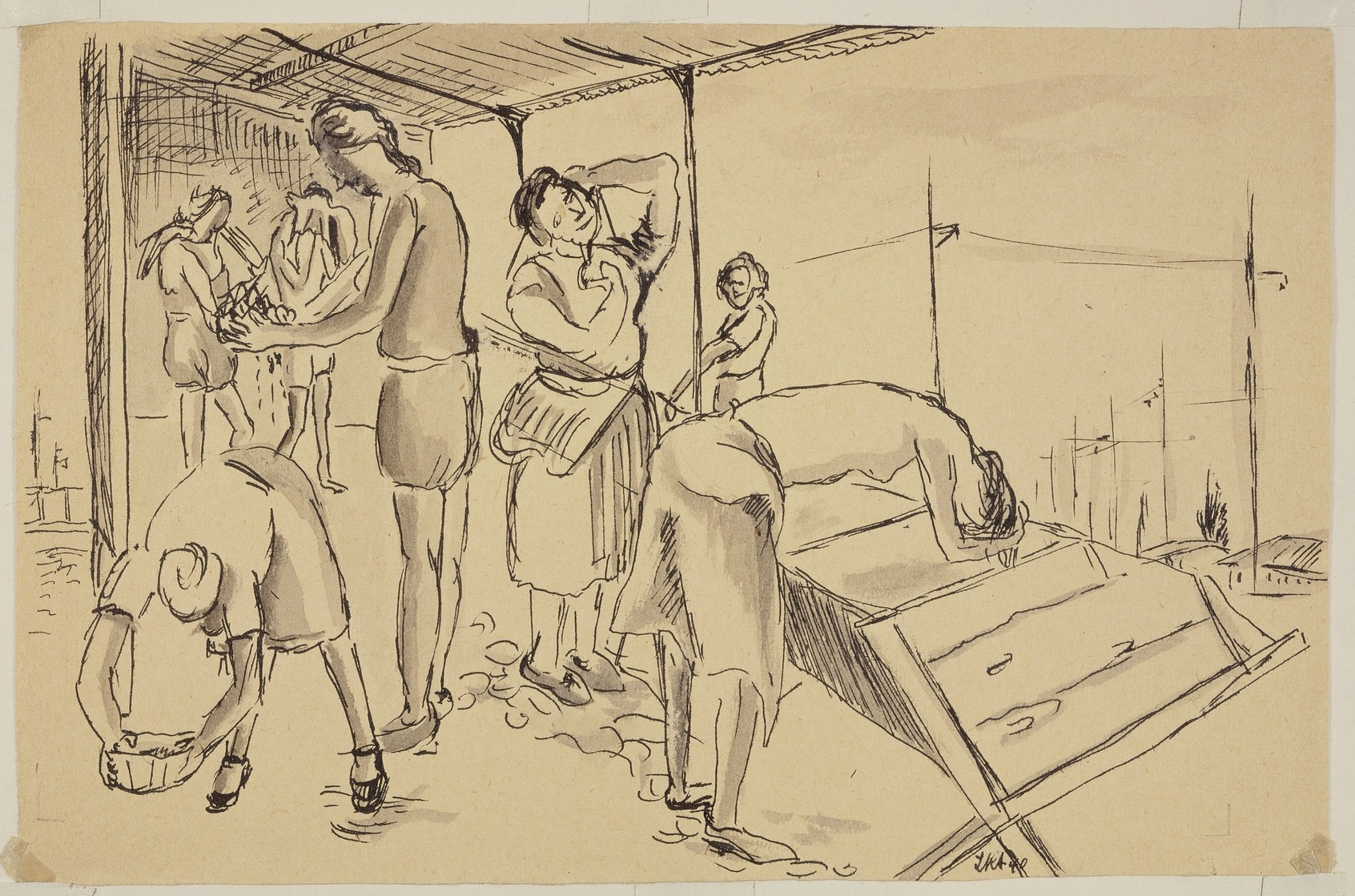}
 \caption{Prisoner artwork example by Lili Andrieux: "Women Washing Themselves (Version I)". Image courtesy: \href{https://collections.ushmm.org/search/catalog/pa1092304}{USHMM}}
 \label{fig:fig1}
\end{figure}

Artists like Leo Haas sought to document the horrors of the Holocaust, holding perpetrators accountable while also capturing moments of dignity\cite{yadvashem2024}. After the war, Holocaust survivors continued creating art to honor the victims. For instance, Mia Fendler Immerman, a child survivor, painted portraits of her family members who died during the Holocaust and, by doing so, preserved their memory\cite{beauregard2018}. Lili Andrieux and Esther Lurie, both represented in our collection, focused on female experiences under these conditions. Andrieux – an example is shown in Figure~\ref{fig:fig1} – highlighted the struggles of maintaining femininity in degrading environments \cite{presiado2016}, while Lurie depicted the physical and emotional toll of the camps on women, emphasizing their resilience \cite{presiado2016}. These artworks not only document suffering but also honor the resilience of the human spirit. 

Only a few projects use this rich source of visual testimonies in digital tools to educate about the experiences of victims of Nazi persecution. One notable exception is the Art Dachau project, where artworks by prisoners are used to accompany a written narrative based on multiple written sources \cite{fink2024}. The Horizon Europe MEMORISE project\cite{memorise2024}, which aims to use digital technologies to capture, preserve, and present the memories and testimonies of Holocaust victims to the public in new and engaging formats, also has recognized this gap. On the one hand, the project creates 3D models on the basis of a few prisoner artworks to create immersive experiences that tell stories about concentration camp life\cite{hall2024}. With this paper, on the other hand, MEMORISE explores distant viewing \cite{arnold2019} methods to quantitatively analyze prisoner artwork collections, bearing on recent advancements in machine learning that have created new ways to explore and analyze artworks. The utilization of object detection for identifying, locating, and classifying objects within an image can be combined with visualization techniques to provide both quantitative and qualitative insights into artworks. This can enhance the analysis of the artwork and create a deeper engagement with it. 

Our project specifically focuses on the application of object detection models on Holocaust-related artworks to create an intuitive and interactive interface. This raises the question: is it actually possible to use object detection on artworks and get reliable results? While artworks present unique challenges, such as varying styles, abstract elements, and intricate details, advances in machine learning have made it more possible. By identifying and classifying objects within these artworks and utilizing visualization tools, we aim to create an interface that can uncover and analyze quantitative patterns within the art collection. Furthermore, our goal is to create an interface that not only provides a quantitative analysis of the artworks but also offers a qualitative exploration, enabling a more intuitive and engaging interaction with the art.

\section{Related Work}

Our project relates to previous methods and techniques for object recognition in visual material and the design of visualizations to analyze cultural heritage collections. Some related works combine both areas, inspiring the development of our dashboard for artwork exploration.

\subsection{Object Detection}
Focusing on the advancements in applying deep learning methods with visual arts, Castellano and Vessio \cite{castellano2021} highlight the new opportunities for analyzing and understanding art. These new opportunities enable the development of automatic tools that can make art more accessible. The authors focus on different models for recognizing and localizing objects in artwork, such as R-CNN, Fast R-CNN, Faster RCNN, and YOLO. They define The R-CNN models as region proposal-based methods and the YOLO framework as a regression-based method. These two kinds of methods differ in performance, as region proposal-based methods usually perform better, whereas regression-based methods are faster - but with a lower accuracy. However, cross-depiction remains a problem for all object detection models, as models trained on real photographs need to generalize across different visual representations. The cross-depiction problem refers to the significant drop in performance that object detection models experience when applied to datasets of different visual representations, as presented in ”Cross-depiction problem: Recognition and synthesis of photographs and artwork” by Hall et al. \cite{hall2015}. In practice, this means that object detection models are challenged when attempting to recognize objects across different depictive styles, as the style that the model has been trained on is very realistic. When tasked with detecting, the model might easily detect a cat in a photograph, but it might not be able to do so in a painting. As a way to address this problem, the authors suggest incorporating spatial and structural information when training object detection models.

\subsection{Visualization}
In recent years, the development of visualizations of cultural heritage, particularly through interactive digital museums, has increased due to the continuous digitization of cultural heritage artifacts \cite{windhager2019}. These visualizations enhance public engagement and educational opportunities within the arts. The following works described may not directly address the visualization of Holocaust artwork - or Holocaust-related themes. However, they provide a good foundation for making considerations and decisions during the design process of our interface.
To accommodate users’ complex information needs, Whitelaw \cite{whitelaw2015} proposes “generous interfaces” designed to promote exploration and engagement by providing rich and easy-to-navigate representations of digital collections. These interfaces often use components like graphs, grids, dynamic filters, and color palettes to offer multiple perspectives.
Bludau et al. \cite{bludau2021} focus on visualizations that highlight the relationships between individual items in cultural collections. They developed an interface that provides both an overview of the collection and detailed relational perspectives, such as a vertical timeline, genre-based organization, and a social relations diagram. This approach to visualization allows users to explore items based on metadata, content, and temporal context, which enhances engagement through diverse perspectives.
Crissaff et al. \cite{crissaff2018} developed ARiES (ARt Image Exploration Space), which is a user-centered interface designed to help art historians explore, analyze, and organize large image collections. It includes tools for image manipulation, annotation, grouping, comparison, and metadata exploration, which was designed based on feedback from domain experts. 
Dumas et al. \cite{dumas2014} explored a tangible user interface for interacting with ArtVis, a visualization tool for exploring around 28,000 European artworks. ArtVis offers three components - Explore, Analyze, and Browse. It uses various visualization techniques like map-based views, time sliders, stacked area charts, and fisheye distortion to promote user engagement. 
A recent survey on storytelling for heritage on Nazi persecution \cite{meffert2024} gives an overview of narrative visualizations that have been developed to support users in exploring and understanding complex themes like the Holocaust. One of the reviewed projects is the ARt-tool project “ARt. Dachau Concentration Camp in Drawings and Paintings” \cite{fink2024}. It combines augmented reality and digital art visualization to enhance holocaust remembrance. It links historical drawings and paintings to their actual locations at the Dachau memorial site through an app and a web tour with a 3D model providing a virtual exploration of the site’s history.
All of these projects demonstrate the potential of innovative interfaces to enhance user engagement and exploration in digital cultural collections by integrating diverse perspectives and interaction techniques.

\subsection{Object detection, visualization and Artworks}
Meinecke et al. (2022) \cite{meinecke2022}  present a virtual museum experience that uses interactive visualizations and machine learning to analyze and explore digitized artworks from the extensive WikiArt dataset, containing over 200k images with diverse metadata. By detecting objects within the images, the system establishes relationships and comparisons across artworks, which enables users to filter by artist style and detected objects. This approach enriches user engagement by revealing underlying connections and enhancing understanding of historical and cultural contexts.
The SMKExplorer \cite{meyer2024} is a visualization interface developed by the IT University of Denmark and the SMK - National Gallery of Denmark. It addresses the cross-depiction problem in art exploration using machine learning models like Contrastive language-image pretraining (CLIP) and grounded language-image pretraining (>GLIP). With customized labels from IconClass, the tool allows users to explore over 109,145 detected objects in 6,477 paintings, offering thematic browsing, metadata exploration, and generative AI (DALL-E 2) for creating new images. While users found the interface intuitive and engaging, challenges with mislabeling were noted, emphasizing the need for careful label selection.
Applying state-of-the-art models for object detection may lead to insufficient results when dealing with collections of paintings in styles and themes that deviate from the models’ training materials. To label a collection of medieval art, Meinecke et al. \cite{meinecke2024} developed a semi-automated annotation framework that combines machine learning with human input to increase the quality of object detection results.
The above-mentioned projects illustrate how digital cultural heritage is still evolving as technology evolves to foster a deeper understanding of the arts. They encourage the users to gain insight, understand, and learn more about the artwork in question. Drawing from these experiences, our project aims to create an interactive visualization interface for exploring Holocaust artwork, encouraging learning and reflection on this historical context.

\section{Methodology}
\begin{figure}[tb]
 \centering 
 \includegraphics[width=\columnwidth]{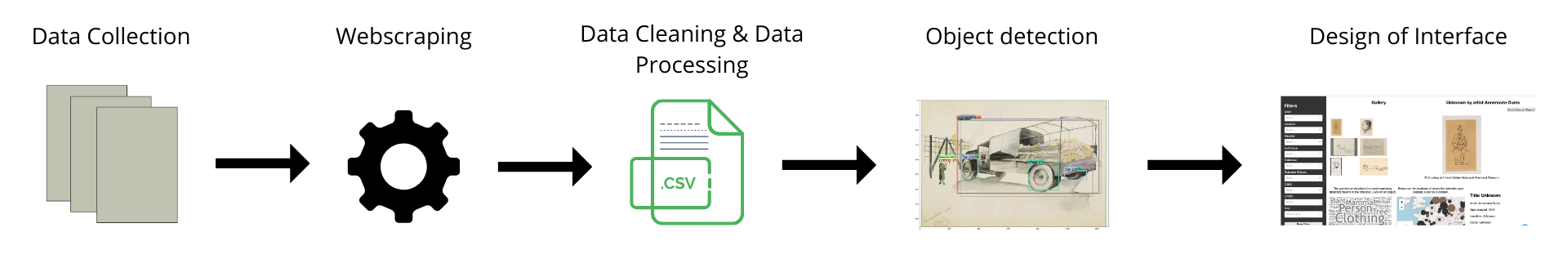}
 \caption{Methodological overview of the project steps from creating the artwork collection to exploring it through the dashboard.}
 \label{fig:fig2}
\end{figure}

Our project included several steps that are depicted in Figure~\ref{fig:fig2}.
\begin{figure*}[tb]
    \centering
    \begin{subfigure}[t]{0.55\textwidth}
        \centering
        \includegraphics[width=\textwidth]{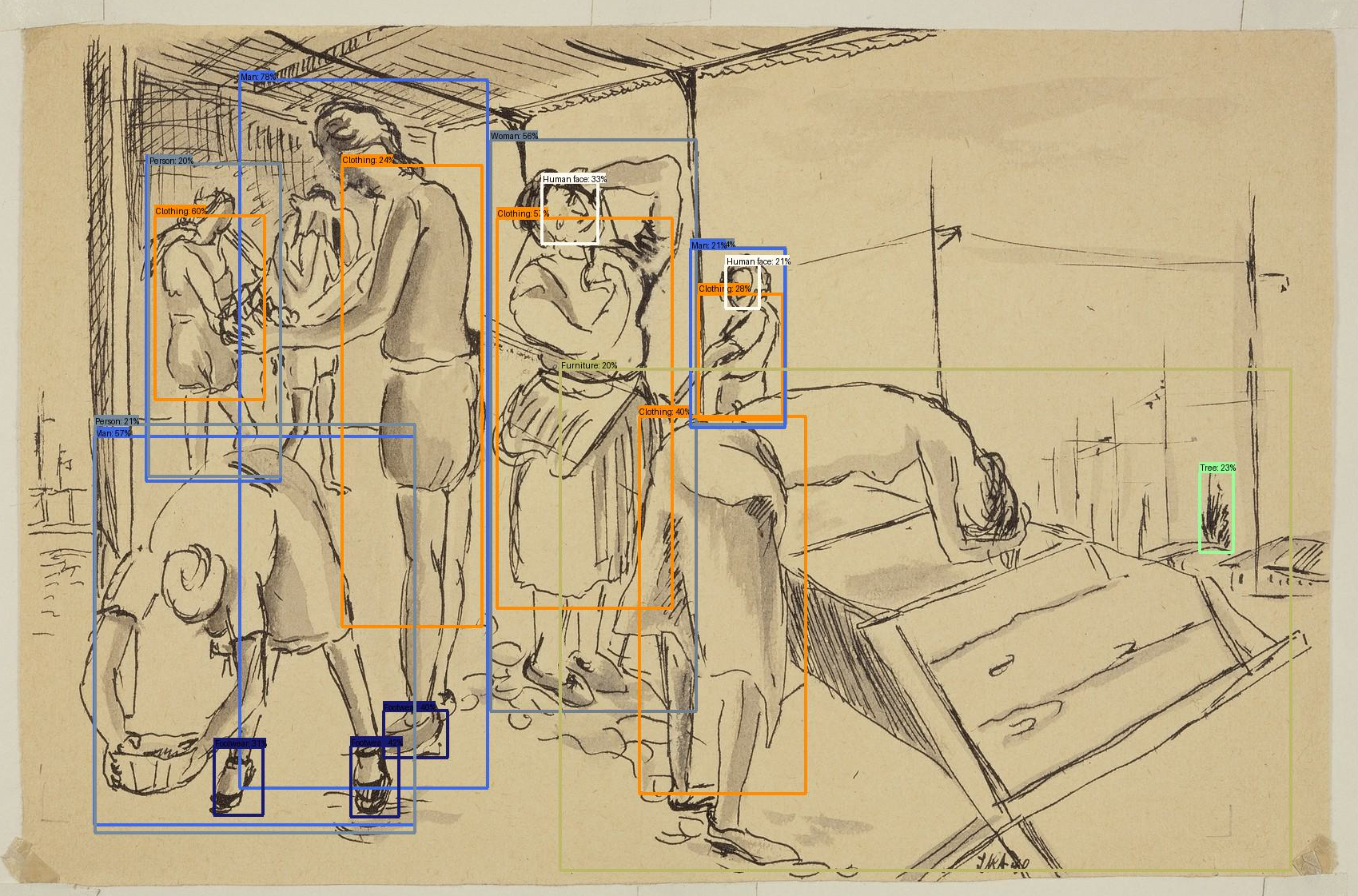}
    \end{subfigure}%
    \hfill
    \begin{subfigure}[t]{0.42\textwidth}
        \centering
        \includegraphics[width=\textwidth]{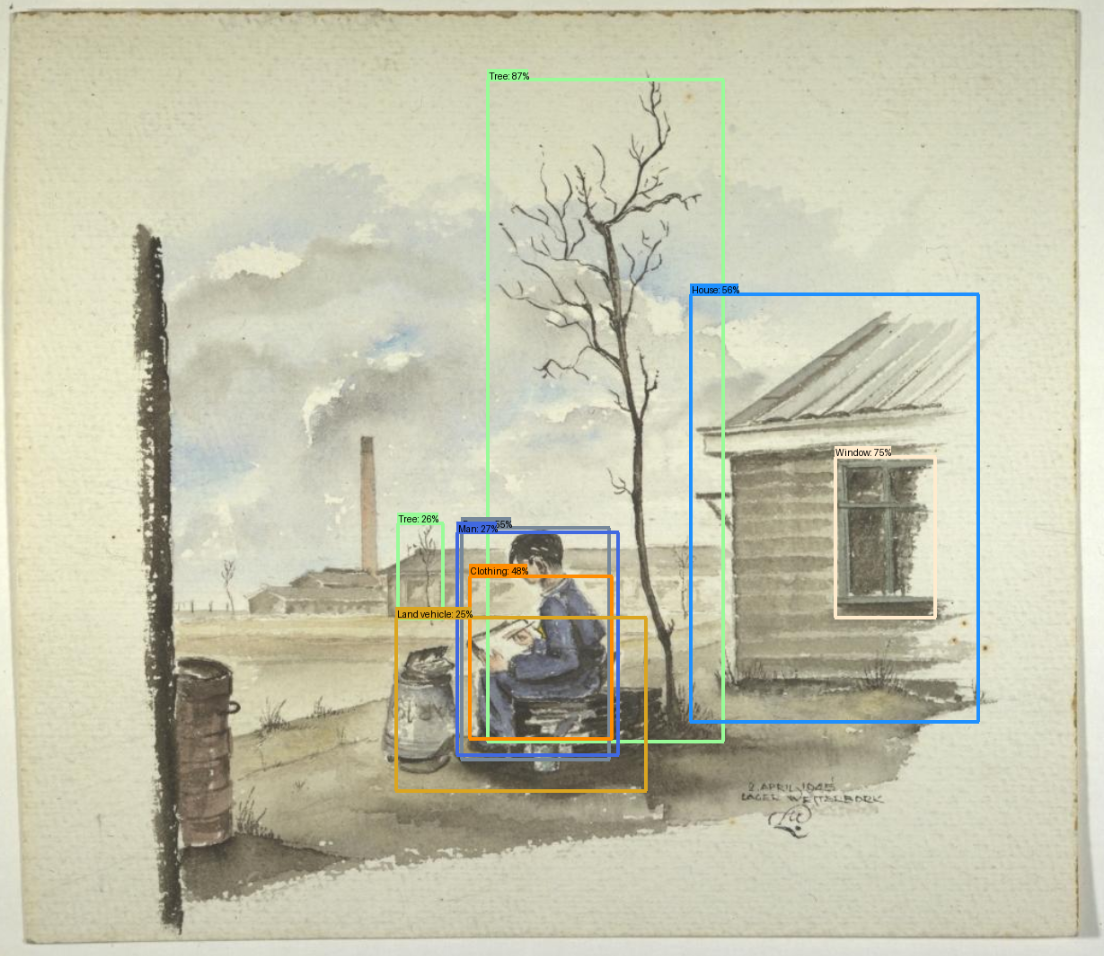}
    \end{subfigure}
    \caption{Detected objects in paintings with a confidence score of at least 20\%. Left: Clothing (orange), Person/Man/Woman (blue), Human Face (white), Tree (green), Furniture (dark green). Right: Painting by Werner Löwenhardt (1945). Image courtesy: \href{https://data.jck.nl/page/aggregation/jhm-museum/M010956}{Jewish Cultural Quarter}. Detected objects: Clothing (orange), Person/Man (blue), Window (beige), Tree (green), House (light blue). Land Vehicle (brown) is a false positive, and a few important objects have not been detected, e.g., the chimney.}
    \label{fig:fig3}
\end{figure*}

\subsection{Data collection, web scraping and cleaning}
The primary objective of this project was to collect an extensive collection of Holocaust artwork, which includes art created before, during, and after the Holocaust. Since there is no single database containing all Holocaust artwork, we utilize web scraping to collect artwork and its metadata from various online sources.
Web scraping involves extracting large amounts of data from websites through scripts, and it is divided into three stages: fetching, extraction, and transformation \cite{vandenbroucke2018}. In the fetching stage, the script accesses the target website using the HTTP protocol. The extraction stage involves retrieving the desired data using HTML parsing libraries and regular expressions \cite{vandenbroucke2018}. Finally, the transformation stage converts the data into a structured format suitable for storage or presentation.
For our project, we collected data from three main databases: the United States Holocaust Memorial Museum (USHMM) \cite{ushmm2024}, Joods Cultureel Kwartier (Joods) \cite{quarter2024}, and NIOD Beeldbank (NIOD) \cite{dutchinstitute2024}. These sources provide diverse collections of Holocaust artwork along with relevant metadata.

After collecting the images and metadata, we performed data cleaning to ensure the quality of the data. This involved manually reviewing images to remove non-artworks and using a script to match images with their corresponding entries in a CSV file. To remove duplicate images that may have been downloaded from multiple sources, we used Gemini 2 \cite{macpaw2024}, a duplicate finder for MacOS, which left us with 1,939 unique images.
We further cleaned the CSV file by reorganizing specific data into the correct columns and manually verifying the results. Additionally, we used the GeoPy library to geocode locations mentioned in the "Location" column, adding latitude and longitude information to visualize the geographical spread of the artwork's creation.

\subsection{Object detection}
\begin{figure*}[h!]
 \centering 
 \includegraphics[width=0.925\textwidth]{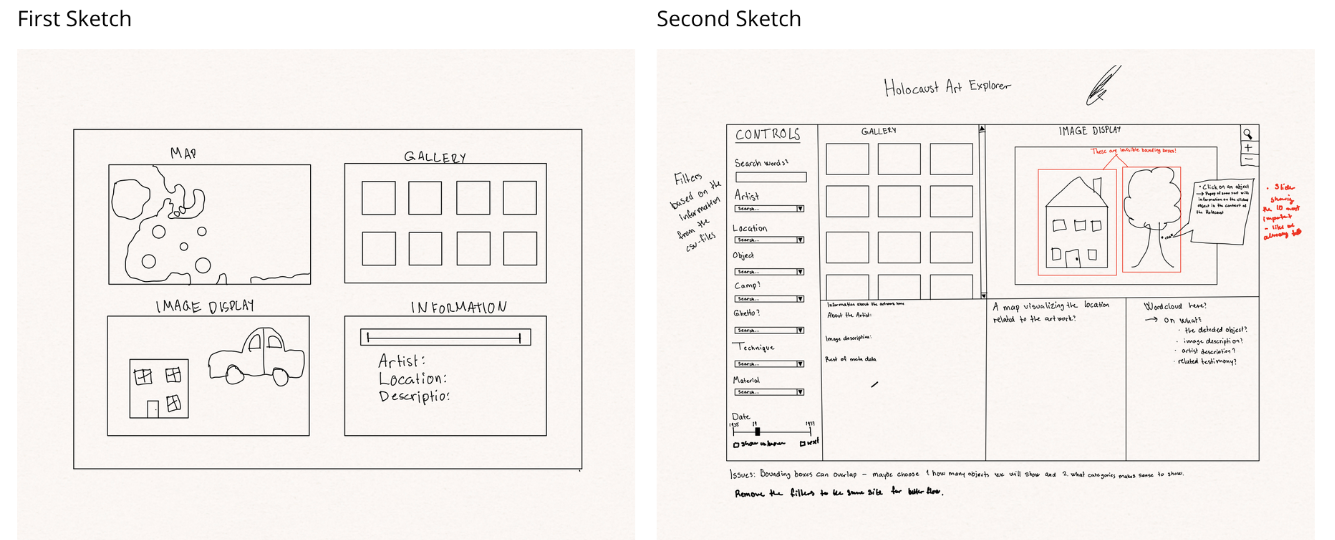}
 \includegraphics[width=0.9\textwidth]{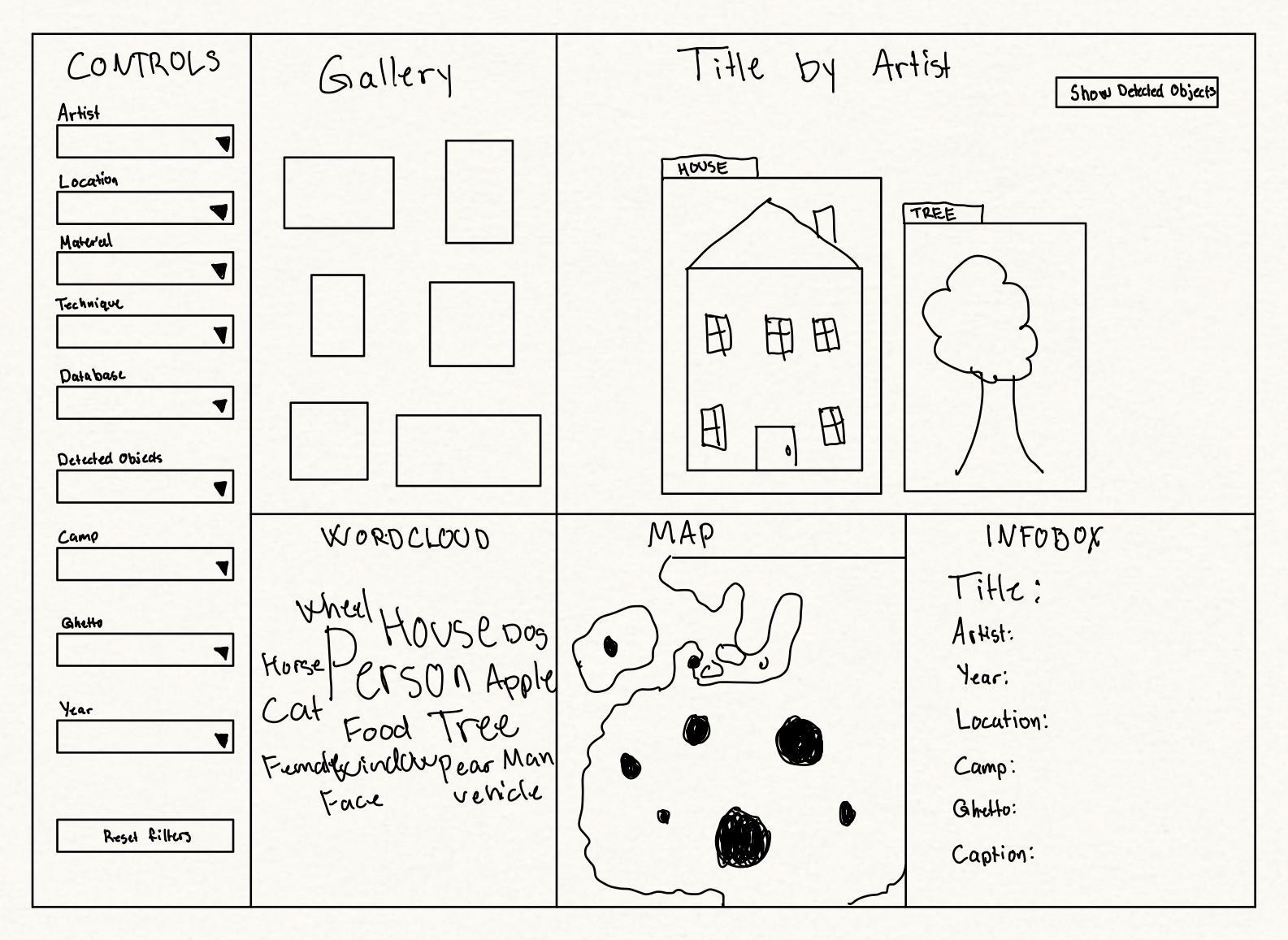}
 \caption{Sketches of iterative Dashboard development in three versions (final version at the bottom).}
 \label{fig:fig5}
\end{figure*}
Object detection is a fundamental task in computer vision that involves identifying and locating multiple objects within an image \cite{michelucci2019}. It goes beyond simple image classification by determining what objects are present and drawing bounding boxes around them, enabling the model to understand spatial relationships within images \cite{michelucci2019}. Deep learning models, particularly convolutional neural networks (CNNs), have significantly advanced object detection but still face challenges known as the cross-depiction problem \cite{hall2015}.

For our project, we chose the Faster R-CNN model \cite{ren2016} from TensorFlow, trained on Open Images V4 with ImageNet pre-trained Inception ResNet V2 as the feature extractor\cite{google2024}. This model was selected for several reasons: it prioritizes accuracy over speed, which aligns with our project's goals \cite{castellano2021}; it detected more object classes compared to the YOLOv8 framework during our tests; and it was easier to adjust according to our requirements. The post-processing settings were configured to display a maximum of 20 bounding boxes to avoid overcrowding and a minimum confidence score of 0.2 to filter low-confidence detections. We stored the detected object information in a CSV file for later use in our interface.
This process resulted in 19,377 detected objects across 1,939 images, providing users with an extensive overview of the artwork contents. While the model can detect up to 100 objects per image, this may lead to some misdetections, such as a violin being detected as a shotgun. These misdetections can have mixed effects: they may spark curiosity and new interpretations but can also cause confusion and detract from the artwork's intended message. By limiting the number of visualized objects to 20 with a confidence score of 20\%, we aimed to balance the richness of exploration with usability. Representative examples are shown in Figure~\ref{fig:fig3}.

\subsection{Dashboard Interface Design}
\begin{figure*}[h!]
 \centering 
 \includegraphics[width=0.49\textwidth]{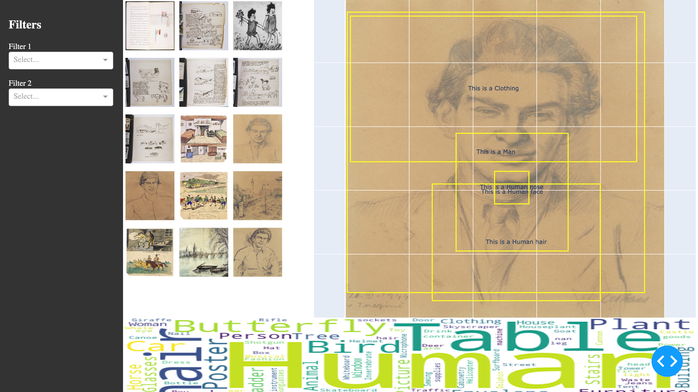}
 \hfill
 \includegraphics[width=0.49\textwidth]{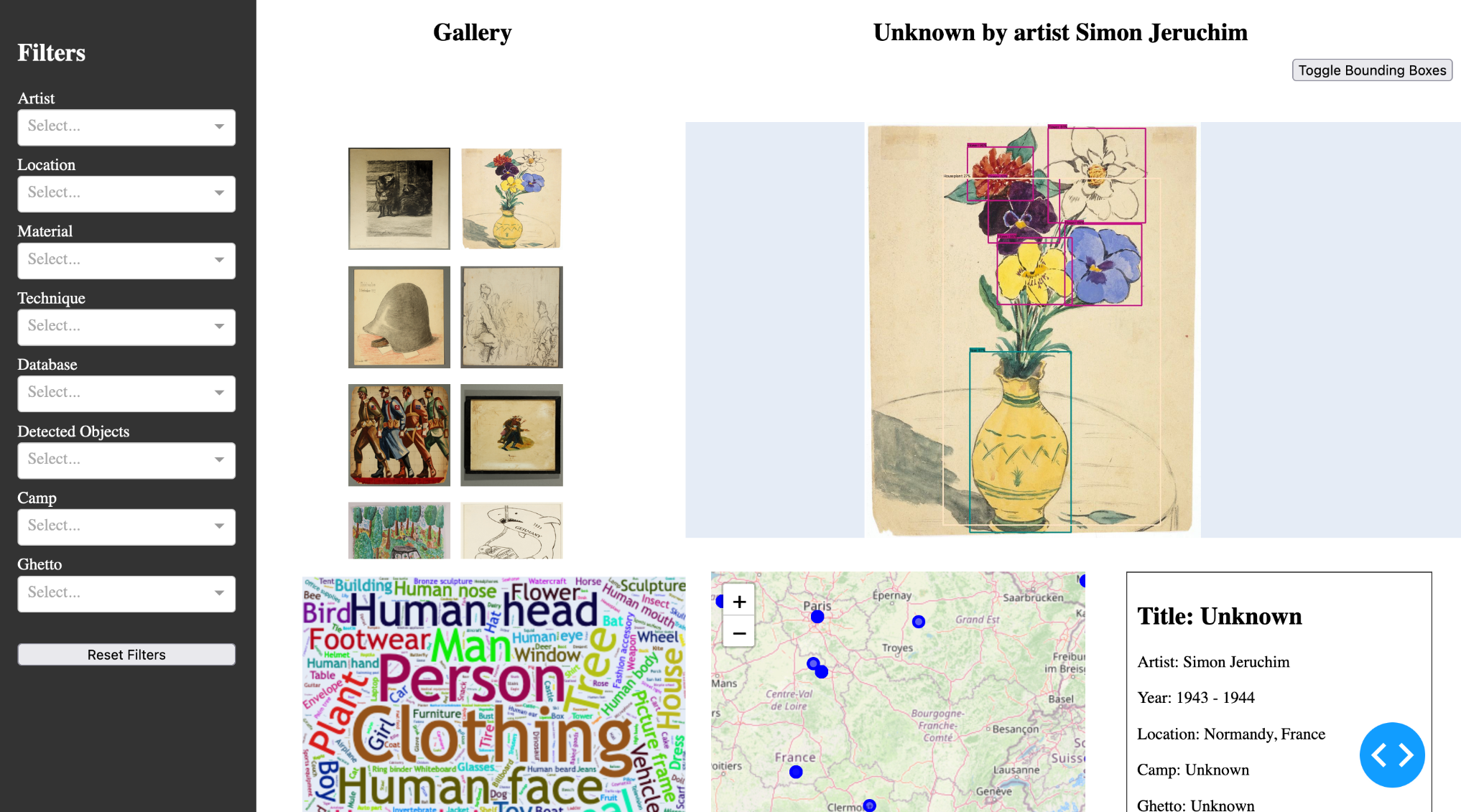}
 
\vspace{5mm}
 
 \includegraphics[width=\textwidth]{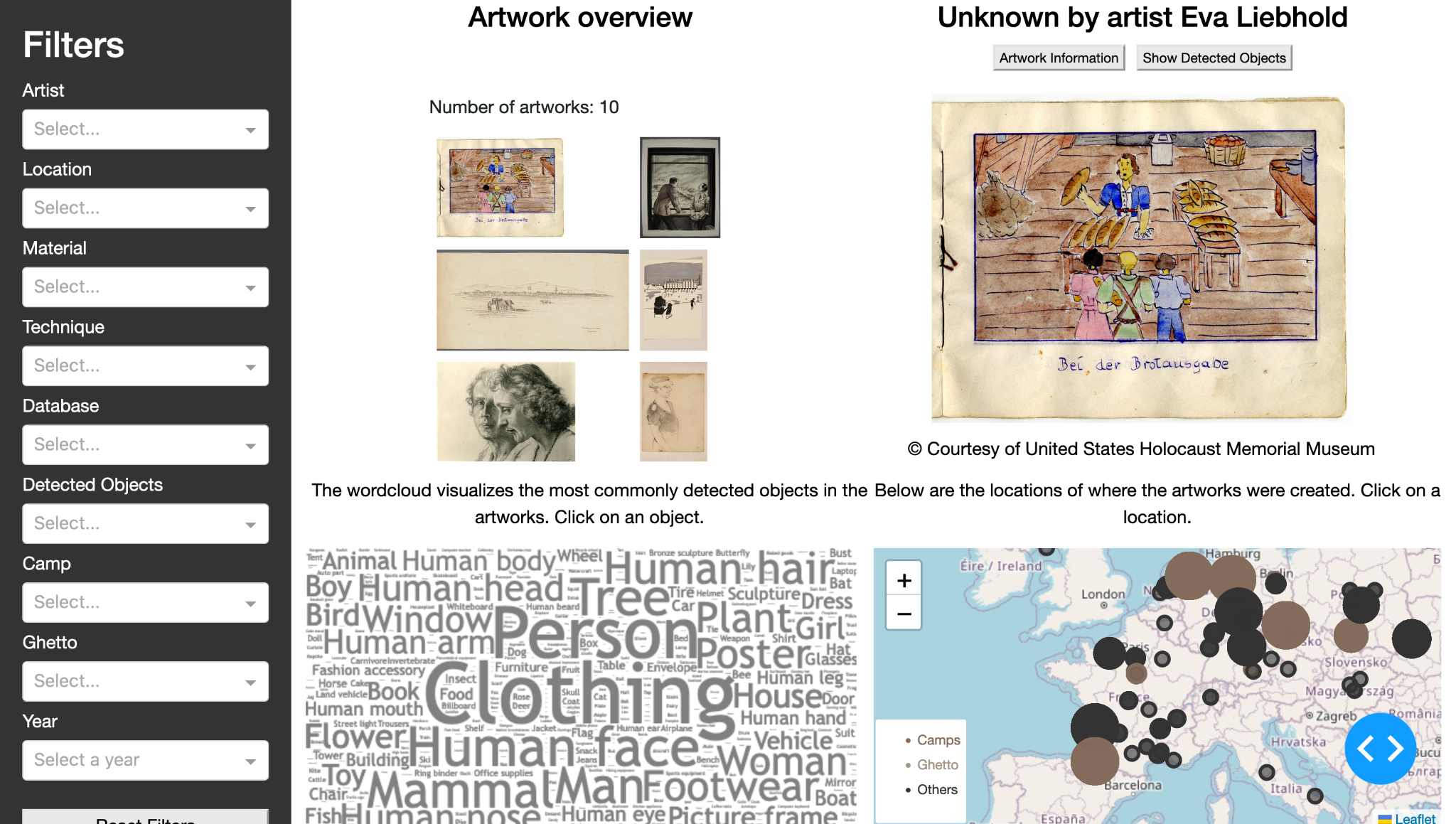}
 \caption{Iterative Dashboard development in three versions (final version at the bottom).}
 \label{fig:fig6}
\end{figure*}
The design and implementation of our interface drew significant inspiration from existing work on visualization, particularly those that highlight effective elements of interactive visualizations [5, 9, 10, 20, 21, 27, 28]. A key objective for our interface is to promote user interaction and engagement with the Holocaust artwork collection, aligning with the ideas of Windhager et al. \cite{windhager2019}. Whitelaw's concept of a "Generous Interface" also influenced our approach, and he advocates for a more exploratory and browsable interface rather than relying on precise search queries \cite{whitelaw2015}. Thus, our interface is designed to provide an interactive and intuitive exploration without requiring specific searches.

To achieve a comprehensive overview of the Holocaust artwork collection, we combined quantitative and qualitative visualization techniques, such as filtering and zooming capabilities. Providing filtering capabilities is also emphasized by Meinecke et al. \cite{meinecke2022} in order to promote user engagement. As is the case by Meinecke et al. \cite{meinecke2022} and Meyer et al. \cite{meyer2024}, we designed our interface to allow users to explore artworks through the detected objects in the holocaust artwork collection.
These design considerations have enabled us to create an interface that not only displays the Holocaust artwork collection but also encourages exploration and engagement. By allowing users to interact with and gain insights from the collection in multiple ways, the interface provides an engaging user experience.

\section{Development of the interface}

Our interface design evolved through several sketches (see Figure~\ref{fig:fig5}), each aimed at enhancing user interaction and exploration of the Holocaust artwork collection. The initial design was a single-page application where all components—such as a geographical map, image display, and gallery overview—were visible without scrolling. Early concepts included making detected objects in images interactive to explore relationships between artworks. However, the initial layout lacked effective filtering options and did not optimize component placement.
The second sketch addressed these issues by adding a sidebar with filters for better control and exploration, allowing users to search through different metadata. This change made the interface more interactive and suited for exploration. The layout was adjusted to place the gallery and image display next to each other, creating a more logical flow. We added a word cloud to visualize the frequency of detected objects and considered how best to use a timeline, eventually grouping it with filters.
In the final sketch, we refined the layout further by grouping the word cloud and map with the filters, as they all serve a similar purpose of updating the gallery. The timeline was replaced with a dropdown filter to save space and focus on the artwork. We also added a feature to toggle bounding boxes around detected objects to avoid visual clutter, giving users more control over their experience. These changes aimed to create an interface that is intuitive, interactive, and encourages engagement with the collection.

\section{Implementation and results}

Just like the design sketches, the dashboard implementation also brought forth three prototypes (see Figure~\ref{fig:fig6}). The final version is designed to allow users to explore the artwork collection through various interactive components. The sidebar enables users to refine their search using filters built with dropdowns that dynamically update the gallery. The gallery displays images in a grid, ensuring proper aspect ratios, and updates based on user interactions with the filters, the word cloud, and the map. 
The image display area shows a selected image, along with the option to toggle detected objects, which are pre-drawn bounding boxes. While the initial plan was to allow object-click interactions for further exploration, this feature is deferred to future work. A word cloud component visualizes frequently detected objects in the artwork collection. The size of the words indicates how often an object is found in the art collection. If you click on a word, a filter is applied that only displays the artworks in which the clicked object was found. The map visualizes the geographical locations of the artworks, enabling users to filter images based on their locations by clicking on the markers. The marker sizes represent the number of images from each location, and the colors represent different types of locations. The infobox provides detailed information about the selected artwork, such as the title, artist, year, and location. When an image is selected, the infobox is updated to display relevant metadata.

Upon entering the interface, users are presented with a randomized gallery of 12 artworks. The first image in the gallery is automatically displayed, and copyright information links to the source database. Below the gallery are the visualizations, including the word cloud and map, which encourage exploration of the collection. Clicking on objects in the word cloud or locations on the map updates the gallery, with a reminder of the user's selection displayed. Sidebar filter selections are shown directly in the dropdowns. The image display also allows toggling between original images and those with detected object overlays. Next to the toggle button is the artwork information button, when clicked it shows an infobox containing information related to the selected artwork.

\section{Evaluation}
\begin{figure}[tb]
 \centering 
 \includegraphics[width=\columnwidth]{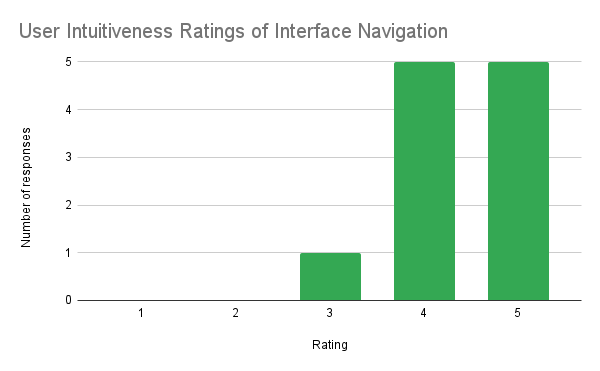}
 \hfill
 \includegraphics[width=\columnwidth]{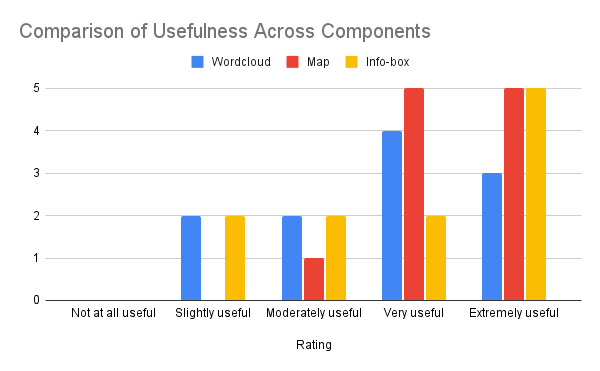}
 \caption{User ratings for intuitiveness of interface navigation and usefulness across interface components.}
 \label{fig:fig7}
\end{figure}

The project was presented to a group of domain experts, who provided feedback after a brief demonstration of the interface. The experts cover the domains of computer science, history, communication, and media, while a few have also done work related to NIOD Beeldbank and the camps Westerbork and Bergen-Belsen. Overall, the experts found the interface intuitive, with 10 out of 11 rating its usability at 4 or 5 out of 5 (see Figure~\ref{fig:fig7} top). This suggests that the interface's layout and design were successful in making it user-friendly and easy to navigate. The experts appreciated the interface's potential for quantitative analysis, particularly its ability to detect objects in artworks, and highlighted the value of the "Show Detected Objects" feature, which allows users to toggle the visibility of colored bounding boxes for better exploration of the art.

Regarding the different components, the map was considered the most useful, with 10 out of 11 experts rating it as "Very useful" or "Extremely useful" (see Figure~\ref{fig:fig7} bottom). The map’s ability to provide a quick overview of artwork locations was praised, with some experts suggesting its integration into larger platforms. The word cloud was also received positively, but some experts suggested improvements, such as allowing users to manipulate the word cloud based on time periods or adding a keyword dropdown list. Others recommended calibrating the word cloud towards the most-used search terms, or even considering its removal. The infobox generated the most polarized feedback, with some experts finding it only moderately useful, while others considered it extremely valuable. None of the visualization components were deemed redundant, but improvements were recommended to enhance their effectiveness.

General suggestions for the interface included refining the layout by adjusting the spacing between components, providing clearer visual structure, and possibly rethinking the placement of elements to improve the presentation. Overall, the feedback was positive and provided insights for future enhancements to the interface. These suggestions will be further explored to elevate the interface to its full potential.

\section{Future Work}
While our prototype still needs improvements, the feedback from domain experts has provided valuable insights.
Some minor work should be done regarding the interface initialization and the overall design of the interface to ensure better flow and enhance how intuitive the interface is. We aimed to add interactivity to the image display by allowing users to click on detected objects to update the gallery. For now, the image display and the visualization of the detected objects are limited to alternating between displaying the artwork with or without pre-drawn boxes around the objects.
Though we could not implement this feature, adding filtering options based on object types or relevance could improve usability. The word cloud was generally well-received, but experts suggested adding filters for detected objects based on year or location. Highlighting detected objects in the word cloud to match selected images would enhance interactivity and exploration.
Object detection revealed 19,377 objects across 1,939 Holocaust artworks, but some misdetections impacted engagement. Training a custom model to focus on Holocaust-related objects like barracks or barbed-wire fences could address this issue. Though we lacked time and resources to train such a model, doing so in the future would improve interactivity and potentially in the future connect artworks with written testimonies.

\section{Conclusion}
We have collected an extensive collection of Holocaust artwork and its metadata through web scraping of the databases, United States Holocaust Memorial Museum, Joods Cultureel Kwartier, and NIOD Beeldbank. This has resulted in a collection of 1,939 pieces of Holocaust art and its metadata. By utilizing the Faster R-CNN object detection model, we have applied object detection to our collection of Holocaust artwork. This process has resulted in 19,377 detected objects within the artwork collection of 1,939 images. While the object detection was relatively successful in regard to the number of detected objects, it has produced a number of misdetections, which highlights the challenge of applying object detection to a varied collection in terms of artistic styles.
We have developed an interactive interface consisting of various visualization components to present our findings through the object detection process and to create an interface for quantitative and qualitative analysis of the Holocaust artworks. This interface consists of a sidebar with numerous filters, a gallery displaying a randomized selection from the (filtered) artwork collection, an image display where users can toggle to show or hide the bounding boxes around the detection objects, a word cloud visualizing the most frequently detected objects, a map visualizing where the artwork has been created, and a togglable infobox providing detailed information about the artworks.
The interface has been designed with the goal of being intuitive, engaging, and useful for both quantitative and qualitative analysis. The interface was presented to a group of domain experts, who provided valuable feedback. The majority of the domain experts found it intuitive and user-friendly, particularly appreciating the toggle function and the map visualization. However, suggestions were made for improvements, especially in relation to the word cloud and the overall layout of the interface.
Overall, this project demonstrates the potential of combining digital technology with Holocaust art to encourage remembrance and education. Positive feedback confirms the interface's value and potential, and we aim to refine it further to benefit experts and the general public for educational purposes.

\acknowledgments{
This project is funded by the European Union’s Horizon Europe research and innovation program under grant agreement No. 101061016. The authors are solely responsible for this work which does not represent the opinion of the European Commission. The European Commission is not responsible for any use that might be made of the information contained in this paper.
}

\bibliographystyle{abbrv-doi}

\bibliography{template}
\end{document}